\documentclass[prb,floatfix,twocolumn,amsmath,amssymb,showpacs]{revtex4}
\usepackage{graphicx}

\begin{document}

\title{Anderson impurity in the one-dimensional Hubbard model on finite
size systems}

\author{S. Costamagna, C. J. Gazza, M. E. Torio, and J. A. Riera}
\affiliation{
Instituto de F\'{\i}sica Rosario, Consejo Nacional de
Investigaciones Cient\'{\i}ficas y T\'ecnicas,\\
Universidad Nacional de Rosario, Rosario, Argentina}

\date{\today}

\begin{abstract}
An Anderson impurity in a Hubbard model on chains with finite
length is studied using the density-matrix renormalization group
(DMRG) technique.  In the first place, we analyzed how the reduction
of electron density from half-filling to quarter-filling affects the
Kondo resonance in the limit of Hubbard repulsion $U=0$. In general,
a weak dependence with the electron density was found for the local
density of states (LDOS) at the impurity except when the impurity,
at half-filling, is close to a mixed valence regime.
Next, in the central part of this paper, we studied the effects
of finite Hubbard interaction on the chain at quarter-filling. Our
main result is that this interaction drives the impurity into a more
defined Kondo regime although accompanied in most cases by a
reduction of the spectral weight of the impurity LDOS. Again, for the
impurity in the mixed valence regime, we observed an interesting
nonmonotonic behavior. We also concluded that the conductance, computed
for a small finite bias applied to the leads, follows the behavior of
the impurity LDOS, as in the case of non-interacting
chains. Finally, we analyzed how the Hubbard interaction and the finite
chain length affect the
spin compensation cloud both at zero and at finite temperature, in this
case using quantum Monte Carlo techniques.

\end{abstract}


\maketitle

\section{Introduction}
\label{intro}

There is an increasing interest in strongly correlated systems containing 
onedimensional (1D) structures, both for their novel properties as well
as for their possible applications. For example, chains and two-leg 
ladders are present in transition metal oxides and in organic conductors. 
Other systems of increasing interest involving strongly correlated
features are nanoscopic devices which have been recently fabricated
and experimentally studied.\cite{goldhaber} In many of these devices,
a droplet of electrons confined in the three spatial directions,
the quantum dot (QD), is connected to metallic leads.
In this case the leads are described by tight-binding models, or in 
general by Fermi liquids. The transport through such devices involve
the formation of a Kondo resonance in the QD. Kondo physics is 
then relevant to understand this kind of devices. \cite{hewson}

It is also possible to connect the QD to conducting leads with 
electron correlations. This is the case of spin valves, one of the
simplest devices used in spintronics,\cite{wolf,pasupathy} where a
QD is attached to ferromagnetic leads.
These leads can be built from materials such as manganese oxides
\cite{manganato_tube} or other compounds containing 
transition metal oxides such as vanadates.\cite{vanadato_tube}
There have also been recent developments in producing devices where
the leads are formed by carbon nanotubes (although ultimately these
leads are connected to metallic terminals) which can be considered
quasi-onedimensional and where electron correlations are presumably
important.\cite{tsukagoshi,biercuk,nygard} In fact, a Luttinger
liquid behavior, typical of correlated 1D systems, has been found in
carbon nanotubes.\cite{bockrath}

The problem of a magnetic impurity or QD in a correlated 
onedimensional system is then an interesting topic both for fundamental
and for technological reasons. From the theoretical point of view, the
most important results for this kind of systems were
obtained by considering a single impurity attached to leads described
by Luttinger liquids. Much less is known when the leads are described
by a Hubbard model. In this case, this model has been treated by
linearizing the dispersion relation and, using bosonization
techniques, the long wavelength behavior is captured. Within this
approach, it was shown that in a system of a magnetic impurity 
connected to Luttinger chains by a Kondo exchange interaction $J$,
the Kondo temperature acquires the power-law expression\cite{leetoner}
$T_K \approx (J )^{2/(1-K_\rho)}$ instead of the conventional exponential
law for noninteracting chains. The Luttinger exponent $K_\rho$ also
determines the conductance through an interacting wire, which is 
given by $G=2 K_\rho e^2/h$.\cite{kanefisher}

In this article we consider the even less studied problem of a 
single-impurity Anderson-Hubbard Hamiltonian on finite-size systems.
Our aim is to provide a detailed microscopic picture of the
competition between the electron correlations on the impurity and the
ones on the leads, and to understand the role of finite length of
the chains.  Although the lengths considered in this study are
still smaller than the physical length of current devices, the study 
on finite-size systems are interesting per se. In fact, it has been
suggested that the Kondo temperature can be very different in finite
chains with respect to the one in infinite leads.\cite{simon-affleck}
It is also expected that the size of nanoelectronic devices will be
further reduced in the near future.
The fundamental problem of the competition between the Kondo effect
and the Luttinger liquid starting from a microscopic Hamiltonian 
was recently studied on infinite systems using an approximated 
analytical renormalization group technique.\cite{andergas05} The
problem of an Anderson impurity in a $t-J$ model has been also
addressed within the Bethe {\it Ansatz} formalism.\cite{schlott03}

Although manganese oxides should be described by a generalized 
ferromagnetic Kondo-lattice model\cite{elbio_manganitas}, and 
vanadates or carbon nanotubes by some variants of the Hubbard or $t-J$
models on ladders\cite{nanotube_model,vanadato_model}, in order to
get an insight on the physical behavior, it is necessary to consider
first the simpler case of leads described by a 1D Hubbard model.
Of course, some of the results obtained are relevant to magnetic 
impurities in strongly correlated quasi-1D compounds such as those
mentioned at the beginning. Results for LDOS for example, could
be compared with those obtained by scanning tunneling microscopy
(STM) experiments.\cite{derro}

Numerical techniques are ideally suited to cope with finite systems.
In particular we mainly employ the DMRG technique,\cite{white} which
provides essentially exact results for various real-space properties
on finite chains. These real-space properties of real-space models
can shed light on the functioning of QD devices.\cite{gazza} In this
sense, in this work we will determine the way in which the Kondo
effect is affected by varying the electron density or the Hubbard
repulsion on the leads by computing the electron occupation and the
square of the $z$-component of the spin at the impurity site. More
importantly, from the spin-spin correlation functions we will study
the ``spin compensation cloud", a possible measure of the more
elusive Kondo screening cloud.\cite{gubernatis} Using this technique,
it has been shown that the screening cloud is reduced by electron
correlations on the leads in a system with two Kondo
impurities.\cite{hallberg-egger} We will extend this approach to
the Anderson impurity where we will make a careful study of
finite-size effects. DMRG results for the spin compensation cloud
have been reproduced to a good approximation by using a recently
developed quantum Monte Carlo technique\cite{sandvik}.
This technique also allows us to study this quantity at finite
temperature.

In order to study transport properties, we will study the LDOS at
the impurity, which is related to the conductance in an
essential way. By implementing some recent developments in
DMRG\cite{feiguin,schollwock}, we will compute the conductance
as the response of the system to a finite but small voltage bias
applied to the leads. We will examine this property as the 
Hubbard repulsion $U$ is varied. 
We will also show that the effect of this variation on both the
LDOS at the impurity and the conductance,
depend in turn on the values of the electron interactions on the
impurity. We will explore various types of
impurities, from the mixed valence to Kondo regimes.

The paper is organized as follows. In Section \ref{model} we
describe the model studied and we provide details of the numerical
techniques employed. In Section \ref{varying_density} we study 
the effect of reducing the electron density from half-filling
to quarter-filling for the case of noninteracting (tight-binding)
leads. In Section \ref{varying_U}, we analyze how the Kondo effect 
varies with the Hubbard repulsion $U$ on the leads at quarter-filling.
Finally, the relation between the results obtained and those
in the previous literature, and the relevance of the present study
to real systems or nanoscopic devices, are discussed in Section
\ref{conclusions}.

\section{Model and methods}
\label{model}

In this paper we use the terms ``impurity" and QD indistinctly, and
we refer to the remainder sites of the chain as the ``leads".
We consider a one-dimensional single-impurity Anderson-Hubbard model 
defined by the Hamiltonian:
\begin{eqnarray}
{\cal H} = &-& t \sum_{i=\leq -2,\sigma} (c^{\dagger}_{i \sigma}
c_{i+1 \sigma} + H.c. ) + U \sum_{i=\leq -1} n_{i, \uparrow}
n_{i, \downarrow} \nonumber \\
&-& t \sum_{i=\geq 1,\sigma} (c^{\dagger}_{i \sigma} c_{i+1
\sigma}+ H.c. ) + U \sum_{i=\geq 1} n_{i, \uparrow}
n_{i, \downarrow} \nonumber \\
&-& t' \sum_{\sigma} (c^{\dagger}_{-1 \sigma} c_{0 \sigma} +
c^{\dagger}_{0 \sigma} c_{1 \sigma} + H.c. ) \nonumber\\
&+& \epsilon' ~ n_{0} + U'  n_{0, \uparrow} n_{0, \downarrow}
\label{hamilt}
\end{eqnarray}
\noindent where conventional notation was used. The first two terms
correspond to the left lead and the following two terms to the right
lead. The next term is the impurity-lead interaction, and the last
two terms are the on-site energy and Hubbard repulsion at the
impurity site, located at site $0$, respectively.  We adopt $t$ as
the scale of energy. In the case of $U=0$, the Hamiltonian 
(\ref{hamilt}) reduces to the single-impurity Anderson model, and
for $\epsilon'=0$, $U'=U$, $t'=t$, to the Hubbard model.

Hamiltonian (\ref{hamilt}) will be studied mostly using the numerical
technique DMRG\cite{white,review} on finite-size clusters of length
$L$ with open boundary conditions (OBC). This algorithm provides
numerically exact results for static properties at zero temperature
with a precision which depends on the number $M$ of states retained.
Most of the results here reported were obtained for $M=600$, except
otherwise stated, asserting us that the integrated weight of discarded
states are of order $10^{-6}$ in the worst case.  On the other hand, 
results for dynamical properties, such as LDOS, should be taken
qualitatively, as discussed below. The same apply to the results
obtained for the time evolution of the current on the links
connecting the impurity with the leads. In order to assess even-odd
effects most of the results reported below were obtained by using 
even and odd chain lengths. Throughout all this paper,
DMRG calculations will be done in the subspace of total $S_z=0$
($S_z=1/2$) for $L$ even (odd). The QD is located in one of two
central sites of the chain when $L$ is even (in the central site when
$L$ is odd).

We have computed static properties such as the electron occupancy 
on each site, $\langle n_{i,\sigma}\rangle$
($\sigma=\uparrow, \downarrow$) and spin-spin correlations from the QD,
$S(j)=\langle S_0^z~S_j^z\rangle-\langle S_0^z\rangle\langle S_j^z\rangle$.

A very important quantity related to the Kondo effect is the Kondo
screening length, which is somewhat elusive to 
compute.\cite{simon-affleck} A possible measure of this Kondo length
is the length of the ``spin compensation cloud"\cite{gubernatis}
defined as the length $\xi$ such that
\begin{eqnarray}
 \sum_{ j=-\xi /2, j \neq 0}^ {j=\xi /2} S(j) = x S(0),
\label{sccdef}
\end{eqnarray}
\noindent
where $x$ is an arbitrary parameter. In the following we adopt a fixed
value $x=0.9$ in order to compare  $\xi$ for the different cases.

One of the main quantities we have studied is the local density of
states, $\rho(\omega)$, at the QD. In the first place, from this
quantity it would be possible to evaluate the conductance in the
linear response regime (see below). In the second place, recent
advances in STM have made it
possible to directly measure this quantity. Since in our DMRG
calculation the measurements are performed when the two added sites
are at the center of the chain (symmetrical configuration), the QD
then is one of these two sites which are exactly treated. Then, we
adopt the approximation of applying the creation and annihilation
operators at the QD on the ground state vector and then determine
$\rho(\omega)$ following the well-known continued fraction
formalism. A more accurate approach would be, after the
application of each of those creation and annihilation operators,
to run additional sweeps for an enlarged density
matrix.\cite{hallberg} In any case, the truncation of the Hilbert 
space is the essential source of error in DMRG, and to estimate the
precision of our approach we have compared results for various
numbers of retained states $M$, from $M=300$ to 600. We would
also like to stress the fact that the conductance in linear
response is related to the LDOS near $\omega=0$ where the
approximation is more precise.  Our calculation starts to be less
precise as we move away from the chemical potential, since then
higher excited states are involved.

For this kind of systems, specially in connection to nanoscopic devices,
the most interesting properties are those related to transport, in
particular, the conductance $G$. When the chains or leads are described
by a tight-binding model, or in general by a Fermi liquid, various
expressions relating it to the LDOS at the impurity are available in
linear response,
as mentioned before. In addition, the Friedel's sum rule (see 
Eq. (\ref{Gfriedel}) below) allows a simple and precise calculation
of $G$, particularly for numerical techniques. In the case of Hubbard
chains, or in general Luttinger liquid leads, one must resort to the
expression, in linear response,\cite{kanefisher}
\begin{eqnarray}
G=\lim_{\omega \rightarrow 0}
\frac{1}{\hbar L\omega}
\int_{0}^{L} \int dx~dt~e^{i\omega t} \langle J(x,t) J(0,0) \rangle
\label{Gdef}
\end{eqnarray}
where the current is defined as
\begin{eqnarray}
J(x,t) = i \frac{e}{\hbar} t_{x} \sum_{\sigma} \langle \Psi(t)|
(c^{\dagger}_{x+1\sigma} c_{x\sigma} - H. c.)| \Psi(t)\rangle,
\label{current_op}
\end{eqnarray}
and $| \Psi(t)\rangle$ is the ground state of the system at time
$t$.

Although such a calculation can be carried out within DMRG, some
recent developments collectively known as time-dependent
DMRG\cite{feiguin,schollwock} allow to compute directly 
$\langle J(x,t) \rangle$ as
$\langle \Psi(t)| J(x) |\Psi(t)\rangle$. The time-evolution is
triggered by the application of a bias voltage $\Delta V$ between the
left and right leads at $t=0$.\cite{schmitteckert,alhassanieh}
That is, at time $t=0$, a potential $-\Delta V/2 n_i$
($\Delta V/2 n_i$) on the sites on the left (right) leads is switched
on. Then, charge moves from one lead to the other in a non-equilibrium
process. This non-equilibrium process was studied analytically in
Ref. \onlinecite{wingreen}. In finite size systems,
$\langle J(x,t) \rangle$ follows an oscillatory evolution as charge
is bounced between the left and right leads. As the length of the
leads are extended to infinity, the period of this oscillation
becomes infinitely long (the leads become infinite charge reservoirs)
and the usual concept of conductance is recovered. It was shown in
Refs. \onlinecite{schmitteckert,alhassanieh} that even for moderate
size clusters, on each period, $\langle J(x,t) \rangle$ reaches a
plateau that corresponds to the value of the bulk limit.
Hence a measure of the conductance can be obtained as the value of
this plateau or the amplitude of this oscillation, that is:
\begin{eqnarray}
G(\Delta V) = \max_t \frac{\langle J(t)\rangle}{\Delta V}
\label{condJt}
\end{eqnarray}
where
\begin{eqnarray}
J(t) = (J_L(t) + J_R(t))/2
\end{eqnarray}
and $J_L(t)$ ($J_R(t)$) is the current on the link connecting the
impurity site with the left (right) lead. Notice that
Eq. (\ref{condJt}) is valid for an
arbitrary finite bias $\Delta V$. The evolution of the ground state
$|\Psi(t)\rangle$ from $t$ to $t+\Delta t$ is computed with a
{\it static} Runge-Kutta treatment of the time-dependent
Schr\"odinger equation.\cite{cazalilla}
Again in this case, a much precise calculation could be performed by
including $|\Psi(t)\rangle$ at intermediate times $\tau$ as targeted
states in the density matrix and by running additional
sweeps.\cite{feiguin} This approach is of course much more expensive
computationally.

In the calculations reported below we adopted a bias $\Delta V=0.01$
and a time step $\Delta t=0.1$.

Finally, we report finite temperature results obtained with quantum
Monte Carlo simulations using the stochastic series expansion (SSE)
algorithm.\cite{sandvik} In this case, we consider chains with 
periodic boundary conditions (PBC). The use of PBC helps to assess
the effects of the Friedel oscillations which appear with OBC, used
with DMRG, but are absent with PBC. SSE with the loop upgrade is a
very efficient QMC algorithm which allows to reach much lower
temperatures and larger sizes than other QMC algorithms. It also
avoids the problem of extrapolation of the Trotter dimension
common in ``world-line" algorithms. It is also
worth to note that SSE works in the grand canonical ensemble, i.e.,
the chemical potential has to be tuned to give a desired density
{\it on average}. Notice that the ``minus sign problem" which affects
quantum Monte Carlo simulations for fermionic models is absent in
1D with hopping between nearest neighbor sites only since fermion
permutations appear just as a boundary effect.

\section{Noninteracting leads, electron densities away from half-filling}
\label{varying_density}

The main purpose of this work is to study the effects of correlations
on the leads. This study will be performed in the next Section. 
Most theoretical studies on Kondo effect have considered the half-filled
case ($n=1$), where the system is particle-hole symmetric.
Now, since an infinitesimal value of $U$ drives the system into an
insulating state, in order to keep the leads metallic in the presence
of correlations it is necessary to study the system
away from half-filling. In Section \ref{varying_U} we
will work at quarter-filling ($n=0.5$).
In this Section, as a preliminary step, we study the evolution of the 
Kondo effect as the filling is reduced from $n=1$ to $n=0.5$
in the absence of interactions on the lead ($U=0$). In particular we
are interested in following the evolution of what at half-filling is
the Kondo resonance, as the electron density decreases.

Model Eq.(\ref{hamilt}), which as stated above reduces to the 
single-impurity Anderson model, was studied by DMRG for several $L$ 
up to 96 sites, and for fillings $n=1$, 0.875, 0.75, 0.625 and 0.5.
Since, as we said in the Introduction, there is a competition between
the correlations on the leads and the correlations at the impurity
or QD, we consider four sets of interactions on the impurity:
$\{U'=8, t'=0.5\}$, $\{U'=8, t'=1.0\}$, $\{U'=4, t'=0.5\}$, and 
$\{U'=2, t'=0.5\}$. In all cases we consider the symmetric case,
$\epsilon' = -U'/2$. For these sets of parameters, the effective
Kondo coupling $J_{eff}=4t'^2/U'$ takes the values 0.125, 0.5, 0.25,
and 0.5 respectively, although strictly this relationship
is only valid for $U' >> t'$, and hence not much applicable to the
last case. In spite of sharing the same $J_{eff}$, the second and
fourth sets behave quite differently as we will show below.

\begin{figure}
\includegraphics[width=0.42\textwidth]{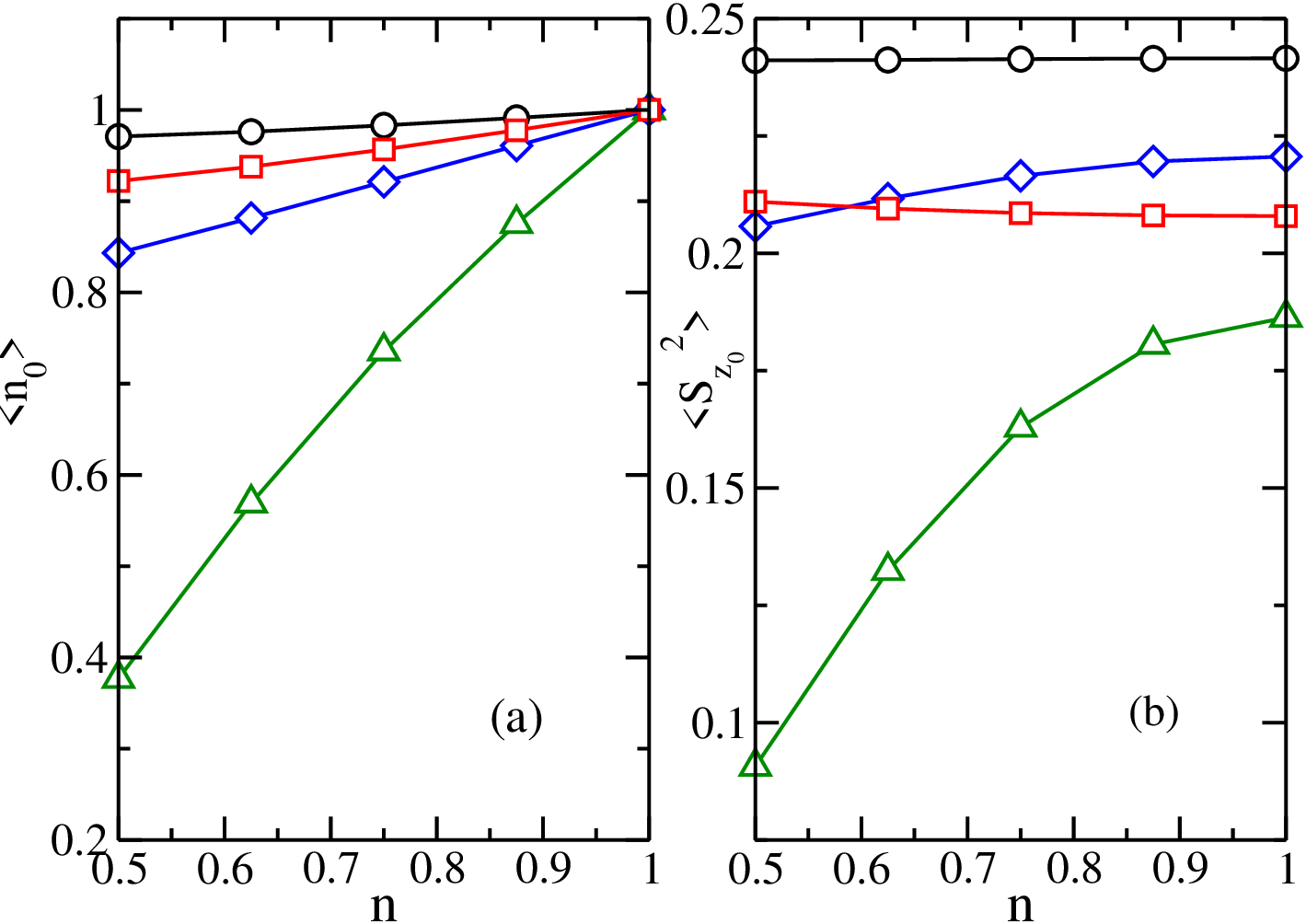}
\caption{(Color online) (a) Electron occupancy and (b)
$\langle S_z^2\rangle$ at the impurity site as a function of the
filling for $U'=8$, $t'=0.5$ (circles), $U'=8$, $t'=1.0$ (squares),
$U'=4$, $t'=0.5$ (diamonds) and $U'=2$, $t'=0.5$ (triangles).
$L=96$}
\label{fig1}
\end{figure}

Let us start by examining the dependence of the electron occupancy
$\langle n_0\rangle$ and $\langle S_{z,0}^2\rangle$ at the impurity site 
with the electron density. This variation depends in turn strongly
on the interactions at the impurity as it can be seen in
Fig.~\ref{fig1}.
If $\langle n_0\rangle=1$, $\langle S_{z_0}^2\rangle$ takes its
maximum value (1/4) when the impurity
is a perfect spin-1/2, corresponding to the full Kondo regime. In
the mixed valence regime, it takes the value 1/12.
For the parameter set $\{U'=8, t'=0.5\}$, the
system is in a well defined Kondo regime, and the dependence with
$n$ is relatively weak. On the opposite case, for
$\{U'=2, t'=0.5\}$ the system is not in a well-defined Kondo regime
at half-filling, and the dependence with the electron filling is
stronger. The reduction of electron density drives the system into
the mixed valence regime. Notice that in this second case, the
occupancy of the impurity is smaller than $n$. Besides, it is well
known that there are Friedel oscillations in $\langle n_i\rangle$
beginning at the ends of the chains with OBC.\cite{whitaffscal} It is
then valid to ask if these oscillations could have some effect on
the electron occupancy at the impurity. In the present case, any
important influence of the Friedel oscillations on the impurity
properties can be ruled out because the results for $L=96$ are virtually 
identical within error bars to those obtained with SSE and using
PBC where these oscillations are absent, at the lowest temperatures
we considered. Of course, the presence of the impurity induces other
density oscillations which overimpose to the Friedel oscillations
induced by the open ends. In fact we have observed in some cases that 
the presence of the impurity changes the period of the open end 
oscillations from $2k_F$ to $4k_F$. This issue is worth to be
examined more thoroughly but it is somewhat outside the scope of the
present work.

\begin{figure}
\includegraphics[width=0.43\textwidth]{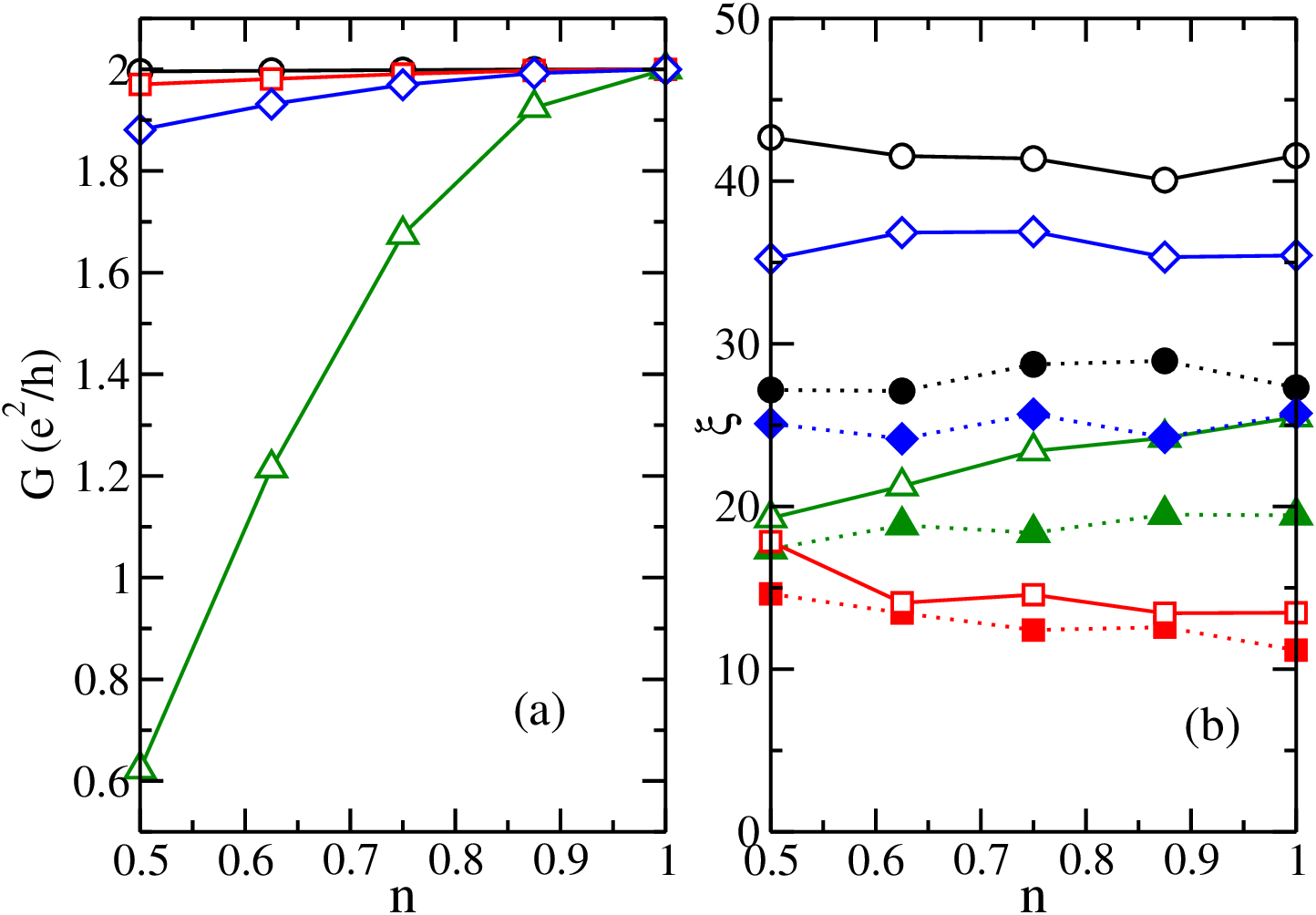}
\caption{(Color online) (a) Conductance obtained from the Friedel's sum
rule, and (b) length of the compensation cloud $\xi$, as a function of
the filling for $U'=8$, $t'=0.5$ (circles), $U'=8$, $t'=1.0$ (squares),
$U'=4$, $t'=0.5$ (diamonds) and $U'=2$, $t'=0.5$ (triangles). $L=64$ 
(full symbols, dot lines), $L=96$ (open symbols, full lines).}
\label{fig2}
\end{figure}

From the values of the electron occupancy it is possible to obtain
the conductance by using the Friedel's sum rule:\cite{hewson,langreth}
\begin{eqnarray}
G_\sigma=sin(\pi n_\sigma)^2
\label{Gfriedel}
\end{eqnarray}
($\sigma=\uparrow, \downarrow$) which is valid when the bandwidth
is much larger than the Kondo coupling. Results for
$G=G_\uparrow +G_\downarrow$ are shown in 
Fig.~\ref{fig2}(a), for the four types of impurities considered.
As expected from the results in Fig.~\ref{fig1}(a), the
conductance decreases by reducing electron density from half-filling.

The length of the spin compensation cloud $\xi$ is depicted in
Fig.~\ref{fig2}(b) for the same four types of impurities. $\xi$
depends on two factors: (i) the effective exchange coupling between
the magnetic impurity and the spins of the conduction electrons, 
and (ii), to a lesser extent, how much defined is the magnetic
character of the impurity or the Kondo regime. The
finite length of the cluster imposes an upper bound to the
compensation length. It has been pointed out that the inability of
the finite system to accommodate the Kondo cloud would modify in turn
the Kondo effect, in particular the Kondo
temperature.\cite{simon-affleck} For the parameter set $U'=8$, 
$t'=0.5$ ($J_{eff}=0.125$) the compensation cloud is large, quite
likely exceeding the system size since $\xi \approx L/2$, and
scaling linearly with the system size. For $U'=4$, $t'=0.5$
($J_{eff}=0.25$), $\xi$ is somewhat smaller but still scales linearly
with $L$. On the other hand, for the $U'=8$, $t'=1$ ($J_{eff}=0.5$)
impurity, where $\langle S_{z,0}^2\rangle \approx 0.21$, the length of
the compensation cloud is much smaller and it has a weak dependence
with $L$. An intermediate situation occurs for the fourth type of
impurity, $U'=2$, $t'=0.5$ which has the same $J_{eff}=0.5$ but with
a poorly defined magnetic character ($\langle S_{z,0}^2\rangle < 0.2$),
leading to a more extended compensation cloud. Notice an overall weak
dependence of $\xi$ with the electron density.

\begin{figure}
\includegraphics[width=0.42\textwidth]{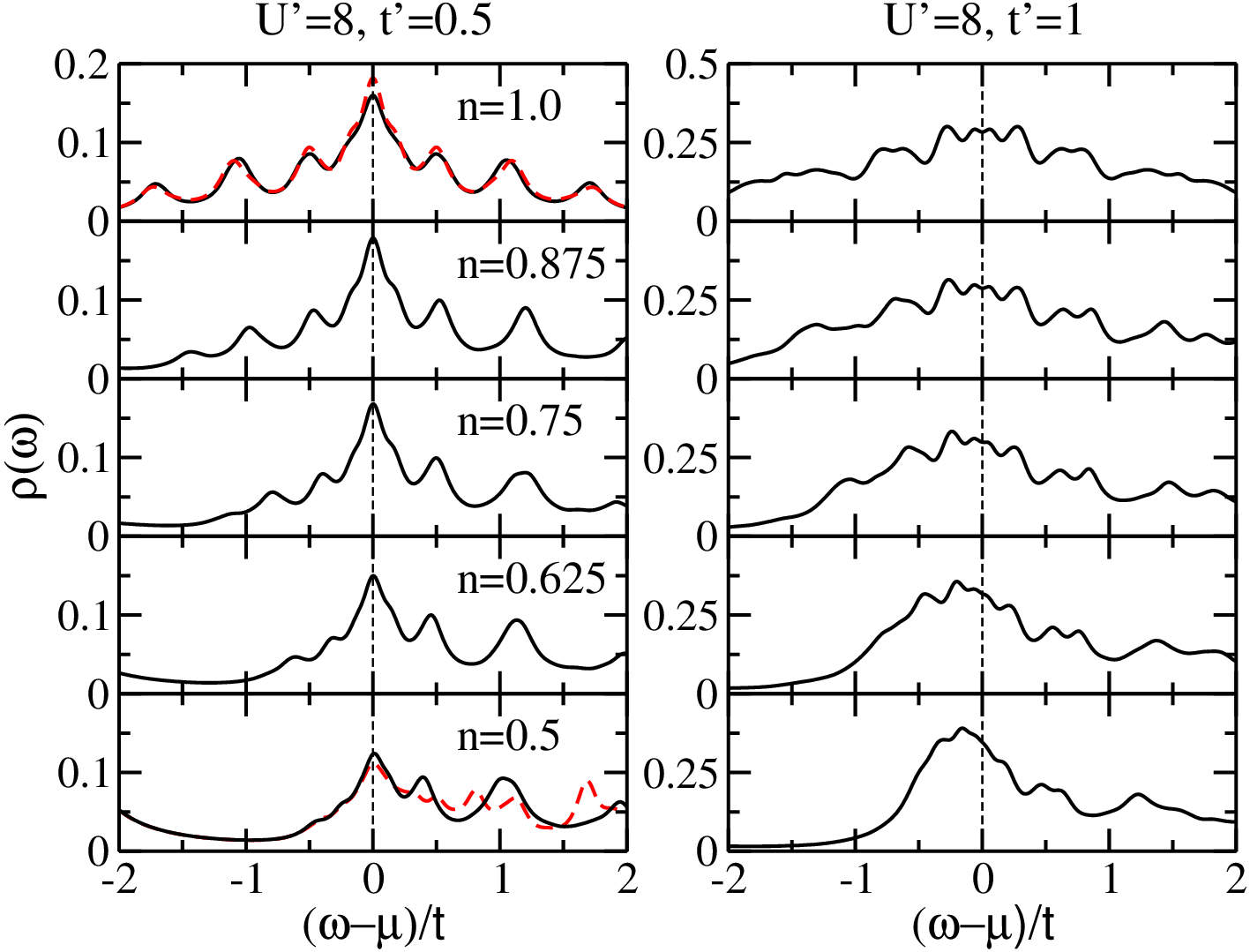}
\vspace{0.6cm}

\includegraphics[width=0.42\textwidth]{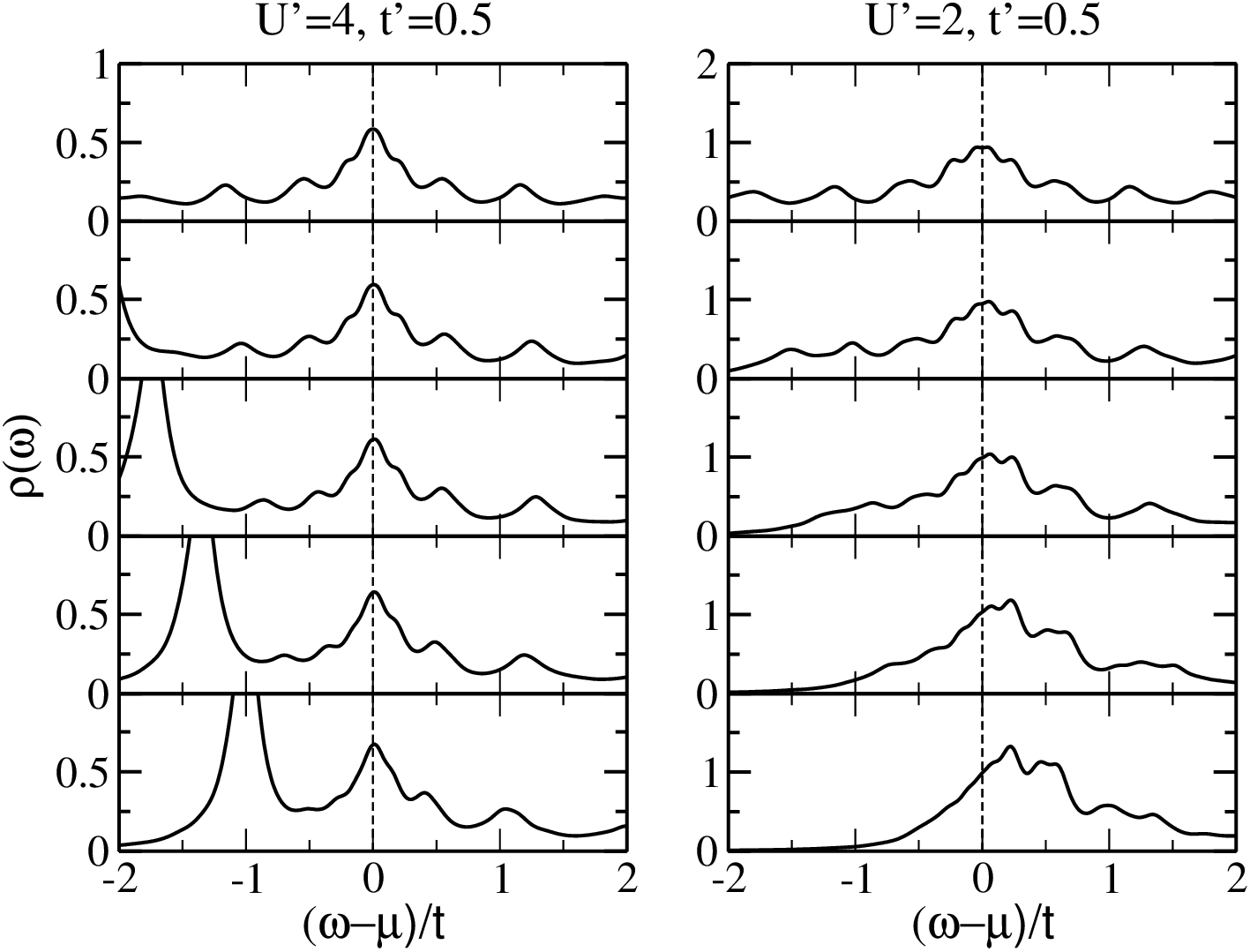}
\caption{(Color online) Evolution of the LDOS at the impurity site as the
electron density is reduced from half-filling to quarter-filling. Titles
on each panel show the corresponding impurity parameters. For each panel,
the fillings are $n=1.0$, 0.875, 0.75, 0.625 and 0.5 from top to bottom,
$L=95$, $M=500$. Dashed lines in the top left panel correspond to 
$L=96$, $M=300$}
\label{fig3}
\end{figure}

It is important to understand the behavior of the conductance shown
in Fig.~\ref{fig2}(a) by studying the LDOS at the impurity site,
$\rho(\omega)$. In Fig.~\ref{fig3} we show $\rho(\omega)$ close to
the Fermi level, for the four types of impurities and for several
electron densities from half-filling to quarter-filling.
$\rho(\omega)$ has been shifted to the chemical potential $\mu$
for the sake of comparison, since starting from half-filling, $\mu$
shifts to $\omega < 0$. In this figure and all the following
similar ones, we adopted a Lorentzian broadening of the peaks of
$\delta=0.1$. A measure of the precision of the results presented
can be inferred from the comparison between $L=95$, $M=500$, and
$L=96$, $M=300$. Clearly, the main features are present in both
cases but it can be seen that the ones for $L=95$, $M=500$ are
more consistent as $n$ is reduced.
For the impurity  $U'=8$, $t'=0.5$, the Kondo peak at half-filling
remains at the Fermi level upon reducing the density down to $n=0.5$.
This is consistent with the relative large conductance of this case.
For the impurity $U'=8$, $t'=1$, the peak shifts slightly away from
resonance as the filling is reduced, going into the electron part of
the spectrum ($\omega -\mu < 0$). Also in this case, the behavior
of $\rho(0)$ is correlated with the conductance weakly varying with
the density. For these two impurities, the Coulomb or holon peaks
fall outside of the frequency range adopted in the plot. In the
case of $U'=4$, $t'=0.5$, the Coulomb peak at $\omega \approx -U/2$
shows up as $n$ decreases. The overall behavior of the Kondo peak
is similar to that of the first impurity and this is related to the 
small $J_{eff}$ of these cases. On the other
hand, for the case $U'=2$, $t'=0.5$, the chemical potential shifts
{\it below} the Coulomb peak at $\omega \approx -U/2$ (the Coulomb
peaks are those located at $\omega -\mu \sim \pm 1$ at half-filling).
This behavior is clearly correlated with the important suppression
of the conductance for this impurity observed in Fig.~\ref{fig2}(a).
For the $U'=8$, $t'=1$, and $U'=2$, $t'=0.5$ impurities, the observed
splitting of the central peak at $n=1$ is just a finite size effect;
the split
between the two central peaks, which correspond to electron creation
($\omega -\mu > 0$) and annihilation ($\omega -\mu < 0$), close
when $L\rightarrow \infty$. Notice also the different amplitudes of
the spectral weight at the Fermi level, roughly proportional to
$J_{eff}$.

\section{Interacting leads, quarter-filling}
\label{varying_U}

After examining the effect of varying the filling in a system
with an Anderson impurity in a tight-binding chain, we now turn
to the central issue of this paper, that is the effect of the
presence of a Hubbard on-site potential on the chain. Then,
we consider the full Hamiltonian defined in (\ref{hamilt}), at
quarter-filling, and taking $U=1$, 2, 4 and 8. In the pure
Hubbard chain, the system behaves as a Luttinger liquid, and 
hence, from the point of view of transport properties is a metal.

\begin{figure}
\includegraphics[width=0.42\textwidth]{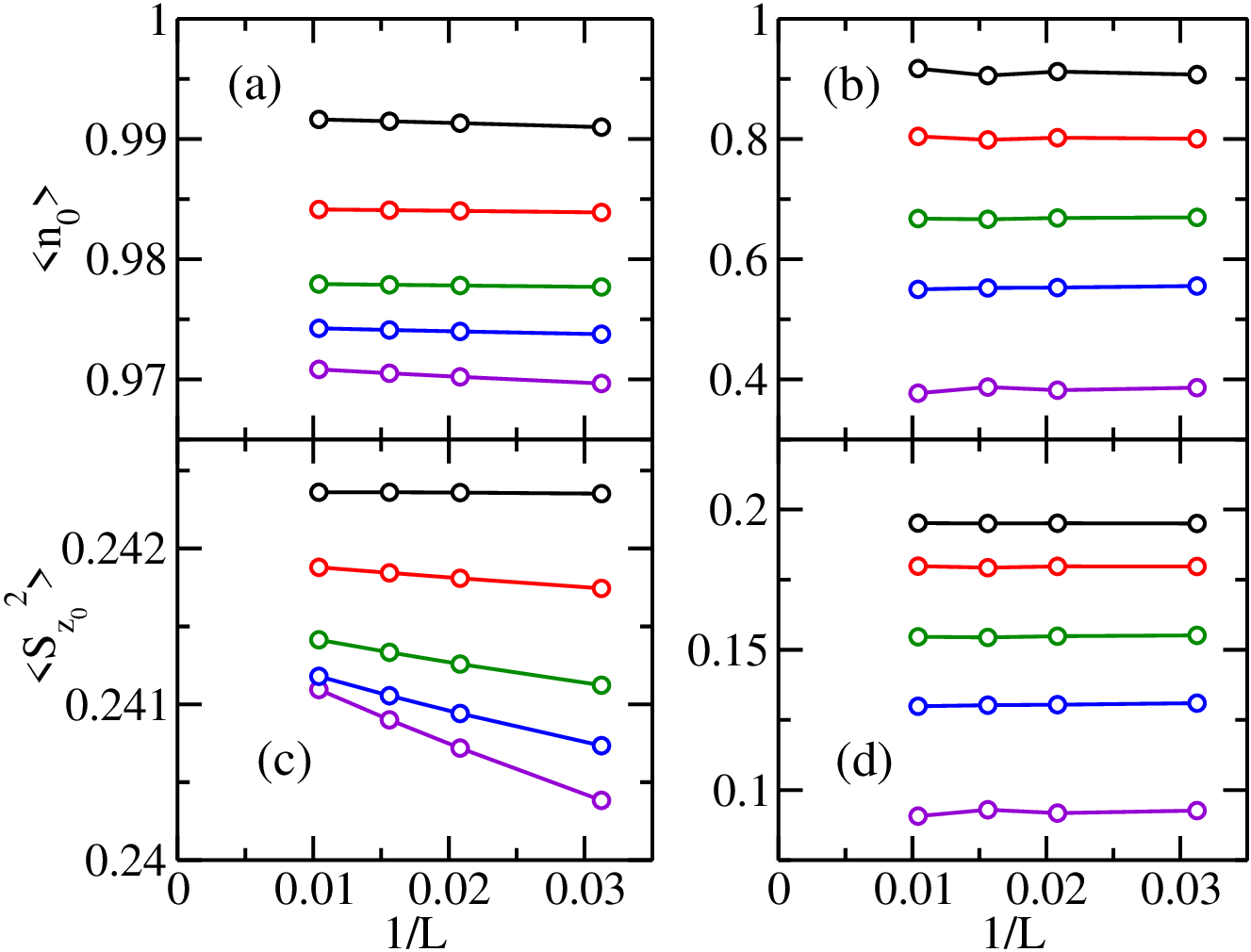}
\caption{(Color online) (a),(b) Electron occupancy and (c),(d)
$\langle S_z^2\rangle$ at the
impurity site as a function of the inverse of the chain length $L$
for $U=0$, 1, 2, 4, and 8, from bottom to top. $U'=8$, $t'=0.5$
(left panel), $U'=2$, $t'=0.5$ (right panel). $n=0.5$.}
\label{fig4}
\vspace{0.6cm}
\end{figure}

We start by looking at finite size
behavior of the electron occupancy and $\langle S_z^2\rangle$ at the
impurity.  Results for the impurities  $U'=8$, $t'=0.5$ and 
$U'=2$, $t'=0.5$ as a function of $1/L$ are shown in Fig.~\ref{fig4}.
For each value of $U$ a weak variation can be observed in
$\langle n_0\rangle$. For the case of $U'=8$, $t'=0.5$ there is a
somewhat larger variation of  $\langle S_{z,0}^2\rangle$ which becomes 
weaker as $U$ is increased.

\begin{figure}
\includegraphics[width=0.43\textwidth]{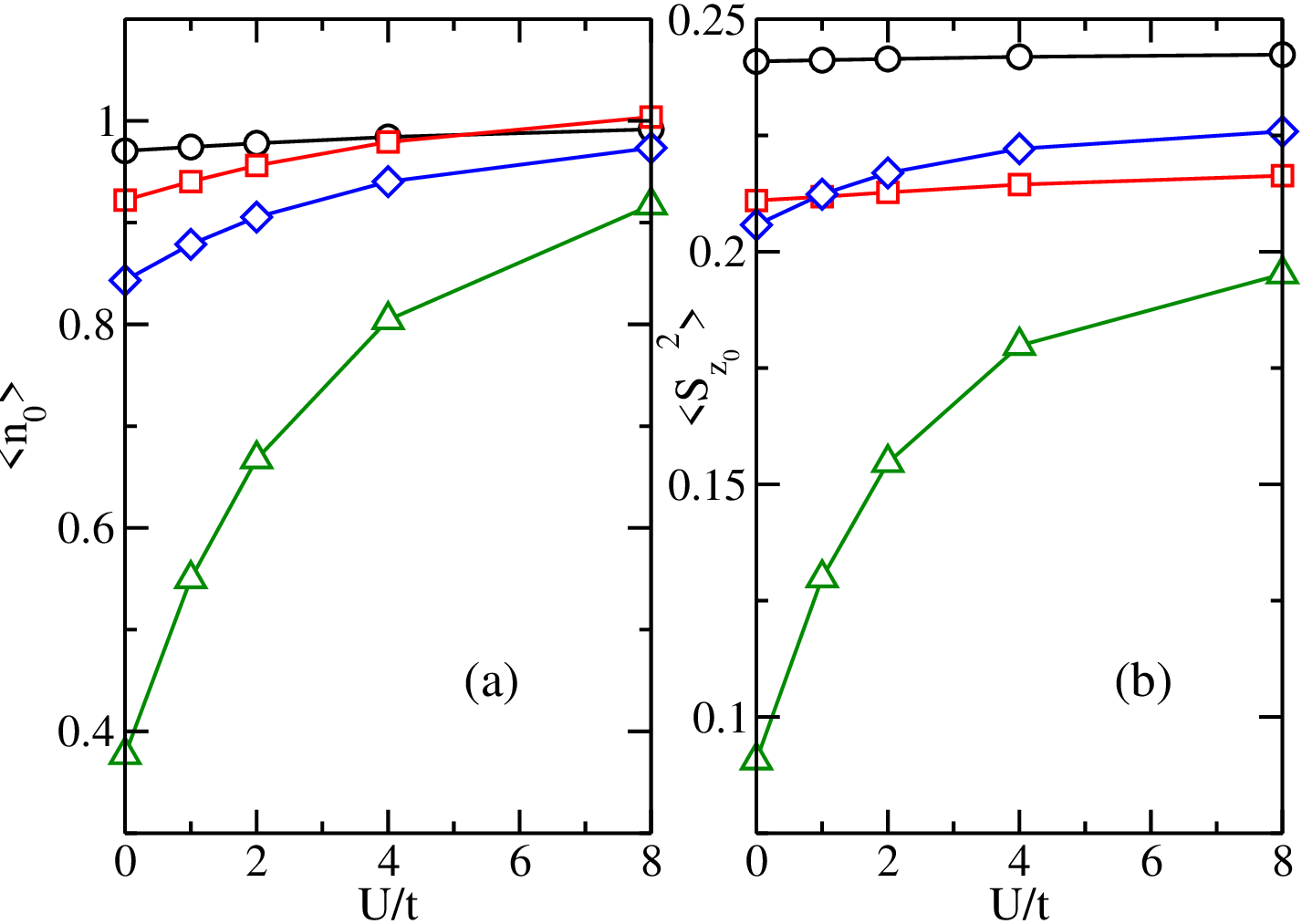}
\caption{(Color online) (a) Electron occupancy and (b)
$\langle S_z^2\rangle$ at the
impurity site as a function of the interaction on the leads $U$ for
$U'=8$, $t'=0.5$ (circles), $U'=8$, $t'=1.0$ (squares),
$U'=4$, $t'=0.5$ (diamonds), and $U'=2$, $t'=0.5$ (triangles). 
$L=96$, $n=0.5$.}
\label{fig5}
\end{figure}

The fact that as $U$ increases, both $\langle n_0\rangle$ and
$\langle S_{z,0}^2\rangle$
increase, is a general behavior observed for all the values of $U'$
and $t'$ that we examined, as can be seen in Fig.~\ref{fig5}. Note
that for $U'=8, t'=1$ in Fig.~\ref{fig5}(a) the system reaches the
limit of the pure Hubbard model at $U=8$, with $\langle n_0\rangle=1$,
while the other cases represent different kind of impurities, with
densities not necessarily equal to 1. This forces the crossing of the
two $U'=8$ curves found near $U=5$. For the cases $U'=2$ and $U'=4$,
for $U>U'$ we can observe a change in the growing behavior of
both $\langle n_0\rangle$ and $\langle S_{z,0}^2\rangle$ curves.
In other words, a Hubbard repulsion on the chains favors a more defined
magnetic character of the impurity. This is particularly apparent for
the impurity with parameters $U'=2$, $t'=0.5$, which most likely is in
a mixed valence state at $U=0$, as discussed in the previous Section.
Again these results are consistent with those obtained with SSE within
error bars. One may conjecture that as $U$ increases, the $4k_F$
electron correlations are enhanced forcing an increased occupation
at the impurity. The enhancement of $\langle n_0\rangle$ and
$\langle S_{z,0}^2\rangle$ with  $U$ occurs also at density
$n=0.75$ at least for this impurity parameters.

\begin{figure}
\includegraphics[width=0.43\textwidth]{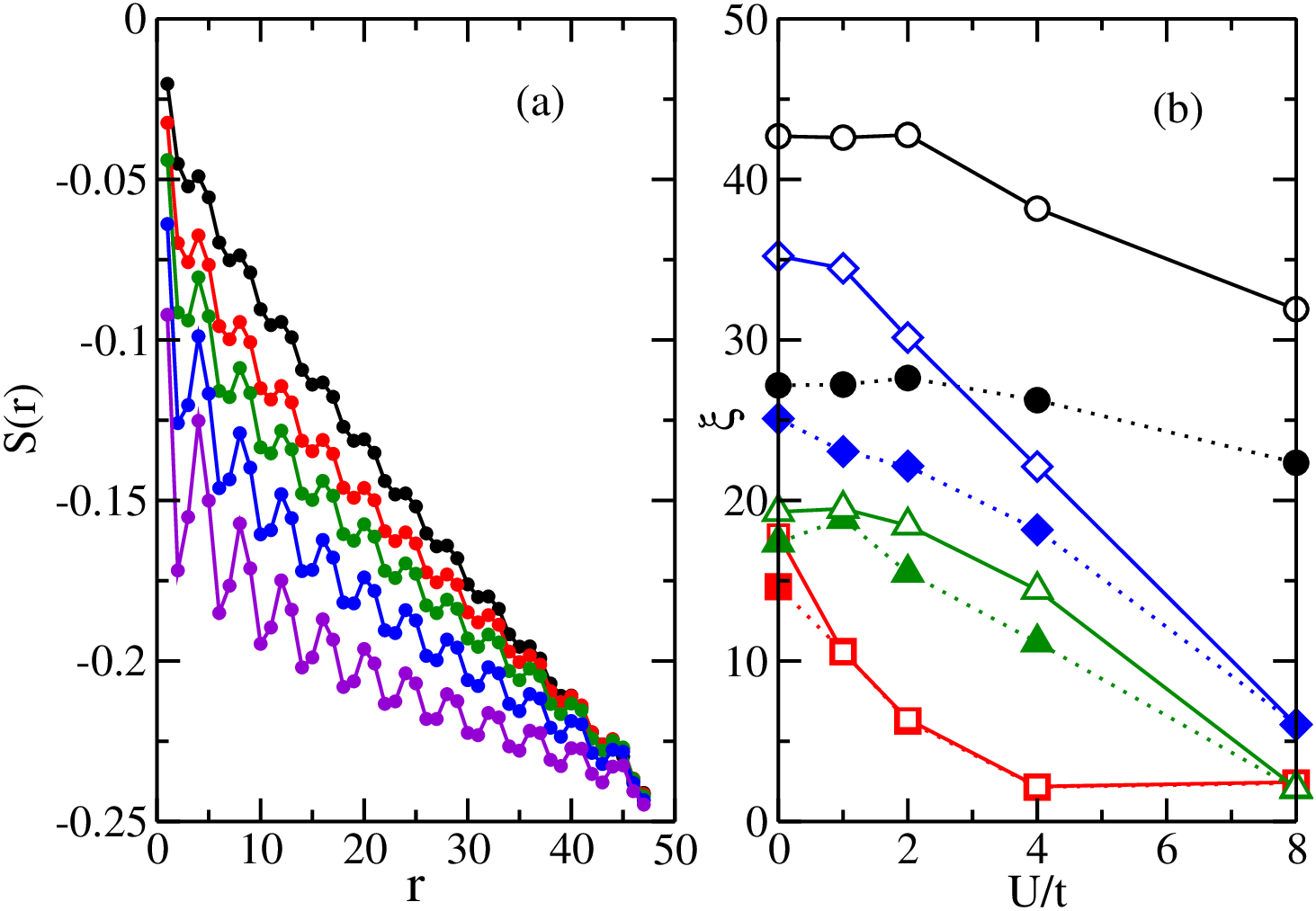}
\caption{(Color online) (a) Integral of the spin-spin correlations as a
function of the
distance from the impurity site (Eq.(\ref{sccdef})), $U'=8$, $t'=0.5$,
and $U=0$, 1, 2, 4, and 8, from top to bottom; $L=96$, $n=0.5$.
(b) Length of the compensation cloud as a function of $U$ for
the four set of interactions at the impurity, $U'=8$, $t'=0.5$,
(circles), $U'=8$, $t'=1.0$ (squares), $U'=4$, $t'=0.5$ (diamonds),
and $U'=2$, $t'=0.5$ (triangles). $L=64$ (open symbols), $L=96$
(filled symbols), $n=0.5$.}
\label{fig6}
\end{figure}

Let us examine now the behavior of the spin compensation cloud when
$U$ is increased. In Fig.~\ref{fig6}(a), the integral of the
spin-spin correlations is plotted as a function of the distance
from the impurity. The typical $2 k_F$ oscillations which at $n=0.5$
has a period equal to 4, are clearly seen. $S(r)$ increases (in 
absolute value) as $U$ increases, implying that the impurity spin
becomes screened at shorter distances. This is precisely the 
behavior of $\xi$, as it can be seen in Fig.~\ref{fig6}(b) for the
four types of impurities considered. For $U'=8$, $t'=0.5$ and $U'=4$,
$t'=0.5$ the reduction of
$\xi$ with $U$ is clear in spite of the fact that the dependence
of $\xi$ with the system size suggests that the bulk limit has not
been reached. On the other hand, for $U'=8$, $t'=1.0$ and for
$U'=2$, $t'=0.5$, results for $L=64$ and 96 are very similar suggesting
that the compensation clouds observed are those of the bulk limit.
This behavior of the spin compensation cloud is correlated with 
the increase of $\langle S_{z,0}^2\rangle$ shown in
Fig.~\ref{fig5}(b). The relative values of $\xi$ for the various
impurities follows in first approximation the values of the Kondo
couplings, as was discussed with respect to Fig.~\ref{fig2}(b).
The overall reduction of $\xi$ with $U$ is consistent with previous
calculations for the two-impurity Kondo coupling\cite{hallberg-egger}
and expected on general grounds due to the increase of the Kondo
temperature with correlations.\cite{leetoner}

\begin{figure}
\includegraphics[width=0.42\textwidth]{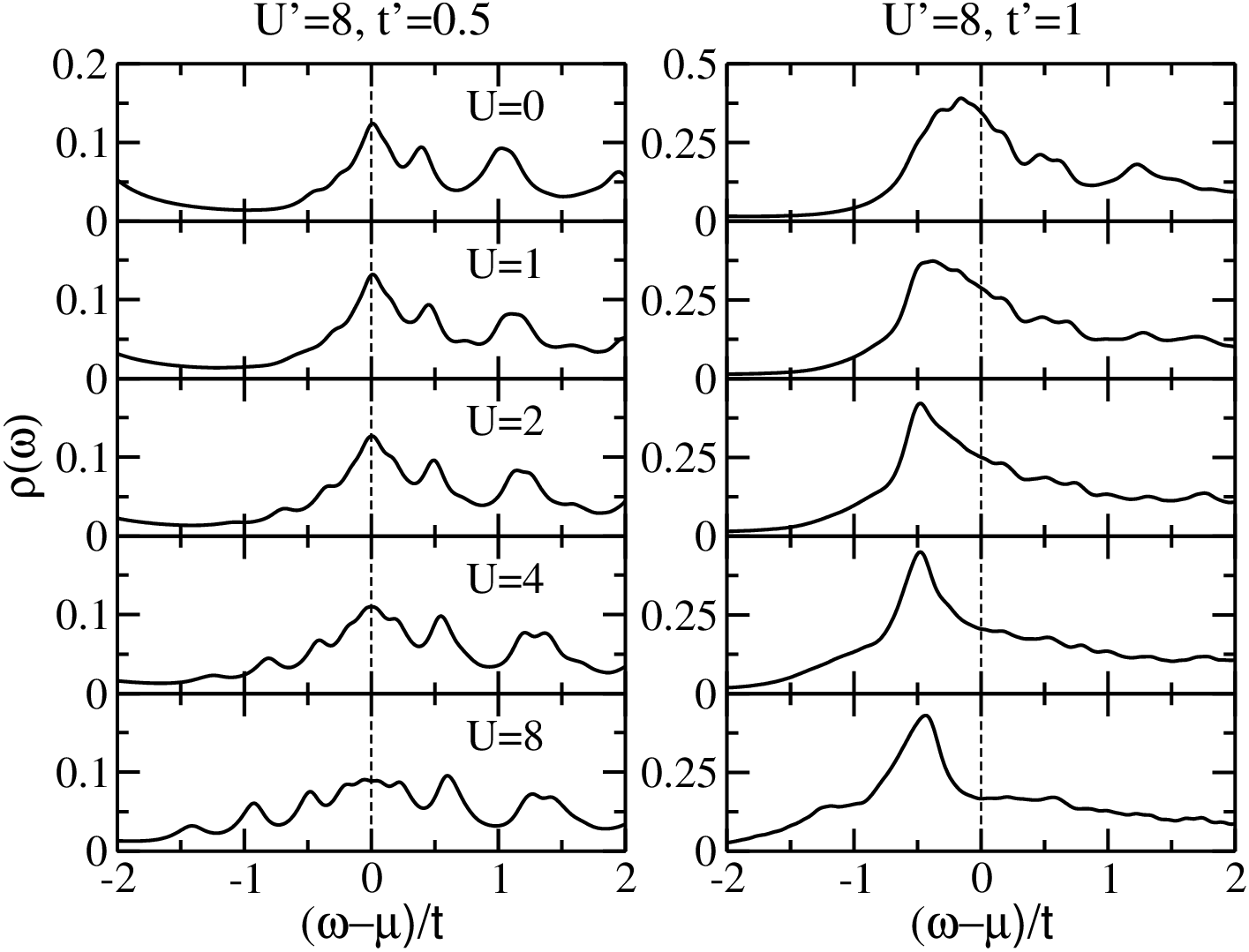}
\vspace{0.8cm}

\includegraphics[width=0.42\textwidth]{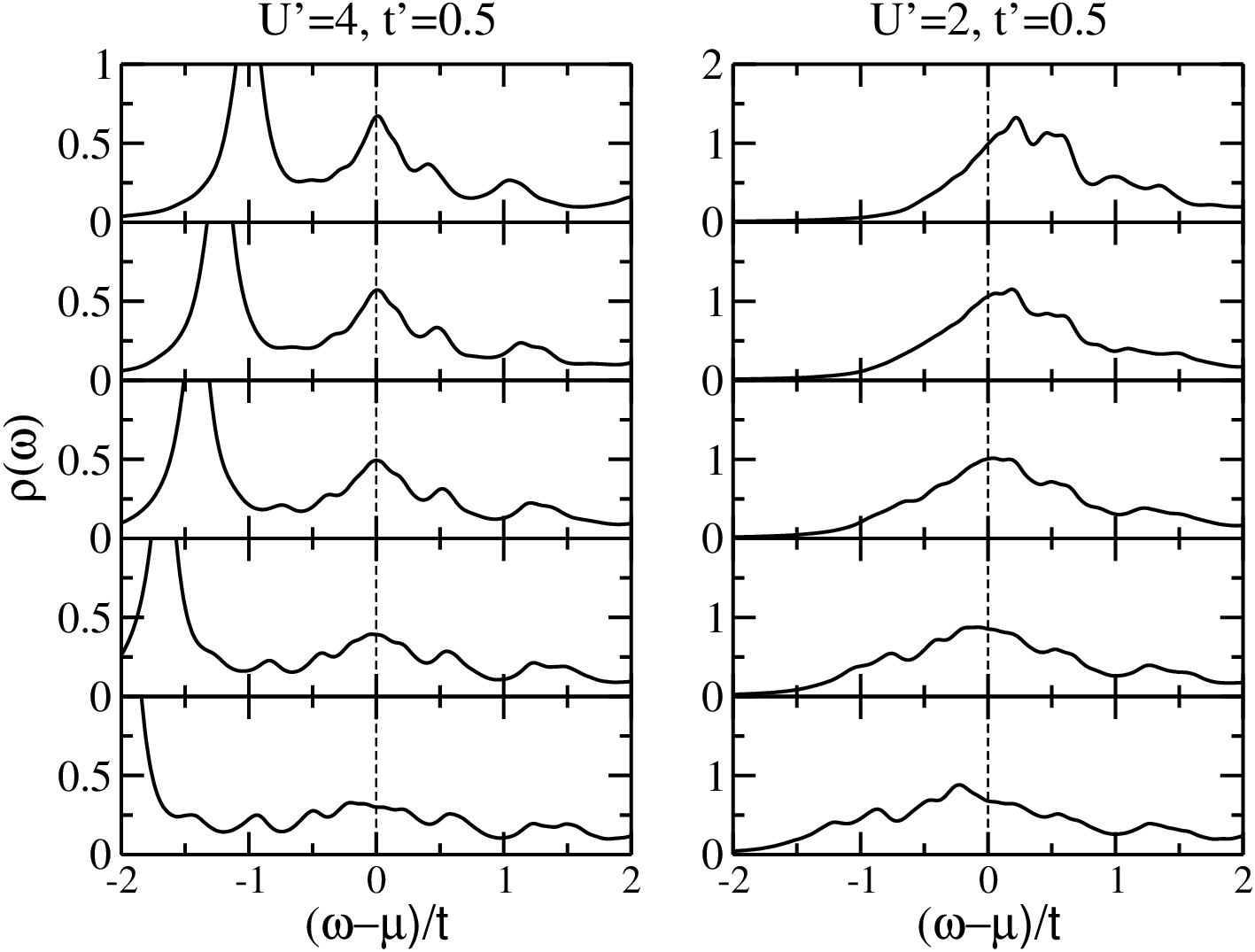}
\caption{Evolution of the LDOS at the impurity site as $U$ in the
leads is increased. Titles on each panel show the corresponding
impurity parameters. For each panel, $U=0.0$, 1.0, 2.0, 4.0 and
8.0 from top to bottom. $L=95$, $n=0.5$.}
\label{fig7}
\end{figure}

The LDOS at the impurity presents also interesting features as $U$
is increased. Let us first consider the impurity $U'=8$, $t'=0.5$
(top left panel Fig.~\ref{fig7}). In this case the peak retains its
resonant character as $U$ increases (only for $U=8$ it appears at a
small negative frequency), and its amplitude is slightly reduced.
For the second type of impurity, $U'=8$, $t'=1.0$ (top right panel
Fig.~\ref{fig7}), there is a pronounced transfer of spectral weight
to the occupied part of the spectrum, and also 
$\rho(\omega -\mu = 0)$ is much reduced. For the impurity $U'=4$, 
$t'=0.5$ (bottom left panel), the remains of the Kondo peak stays
at the chemical potential, which moves towards positive $\omega$.
The same effect of $U$ is observed for $U'=2$, $t'=0.5$ (bottom right
panel Fig.~\ref{fig7}). In this case, the peak located at
$\omega -\mu > 0$ for $U=0$, crosses the Fermi level at $U\sim 4$, 
and finally appears at $\omega -\mu < 0$ for $U=8$, as the spectral
weight is transfered from the unoccupied to the occupied part of
the spectrum. Taking into account the strong enhancement of 
$\langle S_{z,0}^2\rangle$ for these impurity parameters shown 
in Fig.\ref{fig5}(b), one may conjecture that the states close
to the Fermi level that are shifted by increasing $U$ correspond
to magnetic excitations becoming increasingly occupied.

\begin{figure}
\includegraphics[width=0.39\textwidth]{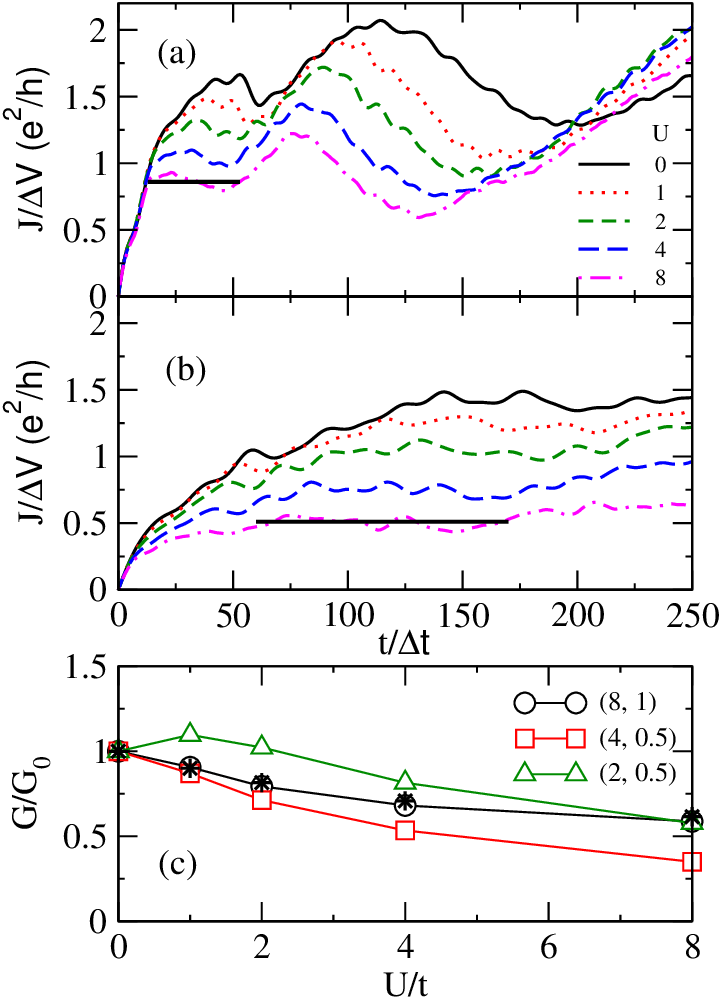}
\caption{(Color online) $J(t)/\Delta_V$ for different values of $U$,
(a) $U'=8$, $t'=1.0$,
(b) $U'=4$, $t'=0.5$, $L=96$, $n=0.5$, $\Delta V=0.01$, $\Delta t=0.1$.
The horizontal lines show the time interval over which the average
of the current is taken to compute the conductance.
(c) Relative conductance (see text) as a function of $U$ for 
different values of the interactions $(U',t')$. The values of 
$K_{\rho}$ at $n=0.5$, extracted from Ref.\onlinecite{schulz},
are shown with stars.}
\label{fig8}
\end{figure}

Our most relevant results regarding the transport properties in a
Hubbard chain in the presence of an Anderson impurity are shown in
Fig.~\ref{fig8}. In Fig.~\ref{fig8}(a,b), we show the average current
$J(t)$ on the links connected to the impurity divided by $\Delta V$
on a $L=96$ chain (results for $L=95$ are virtually identical)
for two sets of impurity interactions. 
As discussed in Section \ref{model},  $J(t)$
has an oscillatory behavior which is typical of these time-evolving
systems.\cite{alhassanieh,meir-wingreen} The period of these 
oscillations for the cases shown in Fig.~\ref{fig8} is approximately
$800 \Delta t$. In this figure, we only show $J(t)$ up to 
$t =250  \Delta t$ because after that the time-evolution becomes 
unstable. In some cases,
(Fig.~\ref{fig8}(a)), these wiggles make the conductance to exceed
the value $2 e^2/h$ indicating that the time-evolution has already
become unstable.

From $J(t)/\Delta V$, we have computed the 
conductance according to Eq.~(\ref{condJt}). Typical time intervals
or ``plateaus" over which the time average was taken are shown in
Fig.~\ref{fig8}(a,b). The values of the conductance $G=G(U)$, for
a given impurity, relative to the value for noninteracting leads,
$G_0=G(U=0)$, are shown in Fig.~\ref{fig8}(c). Although there is
some arbitrariness in defining these plateaus, a qualitatively
clear trend emerges. For the impurities ($U'=8$, $t'=1.0$) and
($U'=4$, $t'=0.5$) $G/G_0$ decreases as U increases. In these
cases, the suppression is consistent with the reduction of
spectral weight at the Fermi level in the impurity LDOS, shown in
Fig.~\ref{fig7}. In these cases, where a Kondo resonance subsists
for finite $U$, and particularly for ($U'=8$, $t'=1.0$), the
suppression of the conductance is consistent with the predicted
relation\cite{kanefisher} $G\sim K_\rho$, as shown in
Fig.~\ref{fig8}(c). On the other hand, for the case of the impurity
parameters  $U'=2$, $t'=0.5$, $G/G_0$ follows a more
complex behavior. From the non-interacting limit
to approximately $U=1$, $G/G_0$ increases and then as $U$ 
increases further, it starts to decrease. This non-monotonous 
behavior again follows the one of the spectral weight at the 
Fermi level in the impurity LDOS (Fig.~\ref{fig7}, bottom right 
panel).

\begin{figure}
\includegraphics[width=0.43\textwidth]{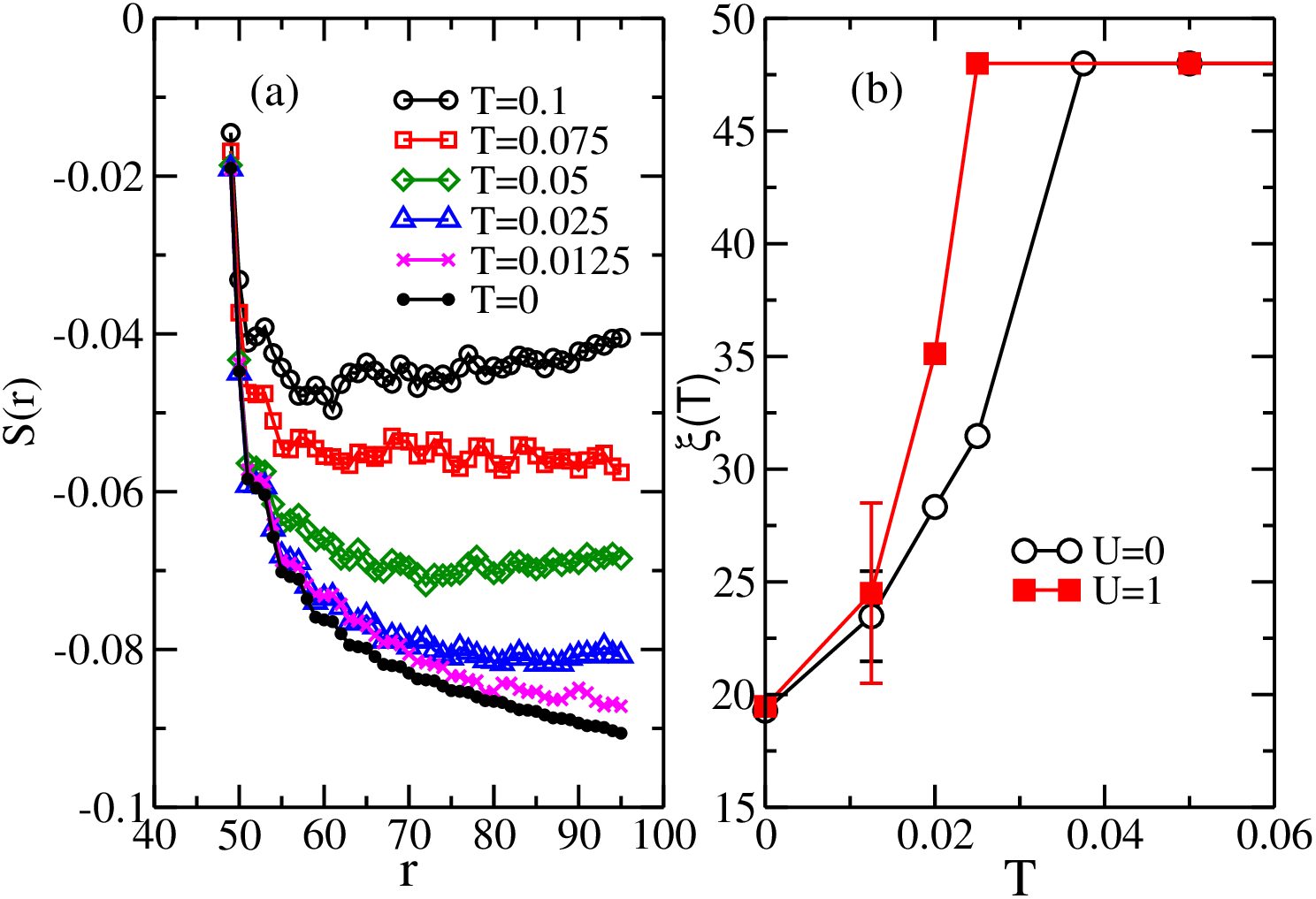}
\caption{(Color online) (a) Evolution of the spin compensation cloud as
a function of distance from the impurity at various temperatures
obtained with SSE, $U'=2$, $t'=0.5$, $L=96$.
(b) Length of the spin compensation cloud as a function of temperature
$U'=2$, $t'=0.5$, $L=96$, $n=0.5$, for $U=0$, and 1. The $T=0$ values
were obtained with DMRG. Error bars are shown for $T=0.0125$. Temperature
in units of $t$.}
\label{fig9}
\end{figure}

Finally, we report a finite temperature study of the spin compensation
cloud. We chose the impurity defined by the couplings $U'=2$, $t'=0.5$.
In this case, since $J_{eff}$ is large (as long as the relation
$J_{eff}= 4t'^2/U'$  is applicable to these parameters), we expect a
relatively small 
length $\xi$ of the spin compensation cloud. In addition, the 
expression for the Kondo temperature for non-interacting leads ($U=0$),
$T_K=t' \sqrt{U'/(2 t)} \exp{(-\pi t U'/({8 t'}^2))}$, although not
much precise for these impurity parameters, gives $T_K \approx 0.02$
which is reasonably accessible by SSE simulations. In
Fig.~\ref{fig9}(a) the spin compensation cloud $S(r)$ is shown at
several temperatures for the case of noninteracting leads, $U=0$. The
temperature is in units of $t$, $k_B=1$.
It is remarkable the convergence of finite temperature results
obtained with SSE to the zero temperature ones obtained with DMRG.
Notice that both techniques are in principle exact but SSE results
correspond to PBC systems while DMRG ones to OBC chains.

The length $\xi$ is shown in Fig.~\ref{fig9}(b). Errors were computed
from the deviation of results obtained in four independent runs.
For $T > T^*$, $T^* \approx 0.375$,  $\xi$ is equal to half
the chain length, implying that the spin compensation cloud exceeds
the limits of the chain. For $T < T^*$, $\xi$ starts to be smaller
than half the chain length indicating that now the cloud is fitting
in the chain. That is to say, for $T < T^*$ the system manages to
compensate the spin of the impurity but this is not possible for
$T > T^*$. If the Kondo temperature is thought as the crossover
temperature around which the impurity spin becomes compensated then
$T^*$ would be a measure of the Kondo temperature in finite systems,
that is, $T^*=T^*(L)$. As mentioned in Section \ref{model}, there is
a certain degree of arbitrariness in the estimation of $\xi$ because
of the cutoff $x$ but what is relevant is to determine how $T^*$
behaves when a parameter of the system is varied. For the case of
$U=1$, $T^*$ reduces its value to $\approx 0.25$. This seems a trend
as the correlations on the leads are increased but results for larger
values of $U$ have larger error bars and so they are not precise
enough to confirm this trend.

\section{Conclusions}
\label{conclusions}

In summary, we have studied the single-impurity Anderson-Hubbard 
model on finite chains with numerical techniques. In the first place,
we analyzed how the reduction of the electron density from 
half-filling to quarter-filling affects the Kondo resonance for
non-interacting leads. Physical quantities defining the magnetic
character of the impurity show in general a weak behavior with the
electron density except when the impurity is, at half-filling,
close to a mixed valence regime. In this case there is a steep
lost of magnetic character by reducing the density. This behavior is
reflected also in the impurity LDOS and in the conductance in linear
response. Next, we turn to the main topic of this 
work which is the study of the effects of a Hubbard interaction
on the leads at quarter-filling. A very clear and interesting result
is that $U$ drives the impurity into a more defined Kondo regime
(Fig.~\ref{fig5}) although accompanied in most cases by a reduction
of the spectral weight of the impurity LDOS. This reduction is
however not general since for an impurity  close to
the mixed valence regime we
observed a nonmonotonic behavior. As we mentioned in the
Introduction, there is a number of strongly correlated materials
which can be or have already been implemented\cite{biercuk} in
nanoscopic devices and where some of the present results for the
impurity LDOS could be contrasted.
The conductance, computed for a small finite bias applied to the
leads, follows the behavior of the impurity LDOS. For the impurities
in a well-defined Kondo regime, the conductance is suppressed by
$U$ in finite chains, in agreement with what is expected for an 
impurity in a Luttinger liquid. Again, for the impurity close to
a mixed valence regime, $G$ increases for small $U$ but eventually
starts to decrease for large values of $U$. This nonmonotonic
behavior indicates a complex interplay between the interactions
on the impurity and the ones on the leads (in addition, of course,
to the impurity-leads interactions).

Finally, we exploit one of the important advantages of the real space
methods we used, that is, the possibility of getting a measure of the
spin screening or compensation of the impurity. The ``spin
compensation cloud" gives an insight on the internal working of the
Kondo effect and it does not depend just on properties at the
impurity site but also on $U$ on the leads. As shown in
Fig.~\ref{fig6},
finite-size effects are more important for small values if $U$.
The proposed determination of the Kondo
temperature in finite systems through the spin compensation cloud, 
has to be thoroughly explored as a function of density and coupling
parameters. We emphasize that this numerical estimation of $T^*$ is
completely general in the sense that it is not limited to 1D
Anderson-Hubbard model but could be used for any Hamiltonian for the
leads, for example the Kondo lattice model, and also not limited to
1D leads, although beyond 1D the ``minus sign problem" makes
difficult the application of QMC methods.  Moreover,  we believe
that the dependence of $T^*$ with the finite size of the system
could be experimentally measured and compared with theoretical
results.

\acknowledgments
This work was supported in part by grant 
PICT 03-12409 (ANPCYT).

\end{document}